\documentclass[twocolumn,showpacs,preprintnumbers,amsmath,amssymb]{revtex4}
\usepackage{amsfonts}
\usepackage{amsmath}
\usepackage{graphicx}
\usepackage{dcolumn}
\usepackage{bm}
\usepackage{overpic}
\usepackage{booktabs}

\begin{document}
\title{Comment on ``Dynamics and Directionality in Complex Networks"}
\author{An Zeng, Ying Fan, Zengru Di\footnote{Email: zdi@bnu.edu.cn}}
 \affiliation{Department of Systems Science, School of Management and Center for Complexity
 Research, Beijing Normal University, Beijing 100875, China}






\maketitle

In a recent letter \cite{PRL103}, authors proposed a residual degree
gradient (RDG) method to enhance networks' synchronizability only by
flipping the direction of the edges without changing the entire
topology and the total weight. In each step, they select the node
with minimum residual degree, and set all of its residual edges
pointing to it (Different from \cite{PRL103}, please note that the
edge direction here is the direction of information flow). The
letter indicates RDG method will finally obtain a network embeds an
oriented spanning tree. They also claim that RDG can enhance the
network synchronization, contrary to the randomly assigned
directional network (RAD).

However, in some cases, the RDG method can not enhance the
synchronizability of the original networks and will actually result
in a directed network with synchronizability
$R=\lambda^{r}_{2}/\lambda^{r}_{N}=0$ \cite{explain}. That means the
RDG method may create more than one ``root node" (the node without
any input). A simple example is given in Fig.1(a). According to the
rule of RDG, node 1 ($k=2$) will be selected first and the two
remaining community will be left disconnected. So two ``root nodes"
(3 and 7) are created respectively. In this case, the RDG network
can never reach complete synchronized state.

We use RDG method in Watts-Strogatz small-world networks
\cite{Nature393} and random scale-free networks \cite{PRE64}.
Specifically, we make sure there is no isolated nodes or communities
in those original networks. When RDG network is with $R=0$, we call
it ``synchronization failure" and the failure rate is defined as the
failure times divided by the total network realization times. The
results are reported in Fig.2. It could be found that the failure is
common when RDG method is used.

To solve the problem of RDG method, we proposed a so-called residual
betweenness gradient (RBG) method. As known to all, degree only
reflects local information. Instead of the degree, we take the edge
betweenness into account, which embeds the global information.
Firstly, we define $s_{i}$ in each node as
$s_{i}=\sum\limits_{j=1}\limits^{N}l_{ij}^{\theta}$, where $l_{ij}$
is the betweenness of edge between $i$ and $j$ and $0 \leq\theta\leq
1$. In fact, when $\theta=0$, $s_{i}=k_{i}$ and RBG is RDG. In each
step, we select the node with minimum $s_{i}$ and set all of its
residual edges pointing to it. If there are multiples of the same
rank $s_{i}$, we choose the node with smaller initial $s_{i}$ first.
Fig.1(b) and Fig.2 show that RBG can solve RDG's problem. And the
detail performance of RBG method will be reported elsewhere.

\begin{figure}
  \center
  \includegraphics[width=5.7cm]{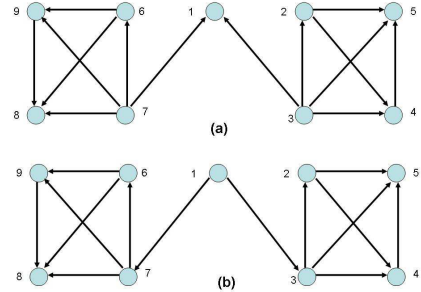}
  \caption{(a)The RDG network from a given undirected network whose original $R=0.039$. The synchronizability
  of this RDG network is $R=0$.(b)The RBG network from the same
  original network when $\theta=1$. Its $R=0.33$.}
\end{figure}

\begin{figure}
  \center
  \includegraphics[width=4.2cm]{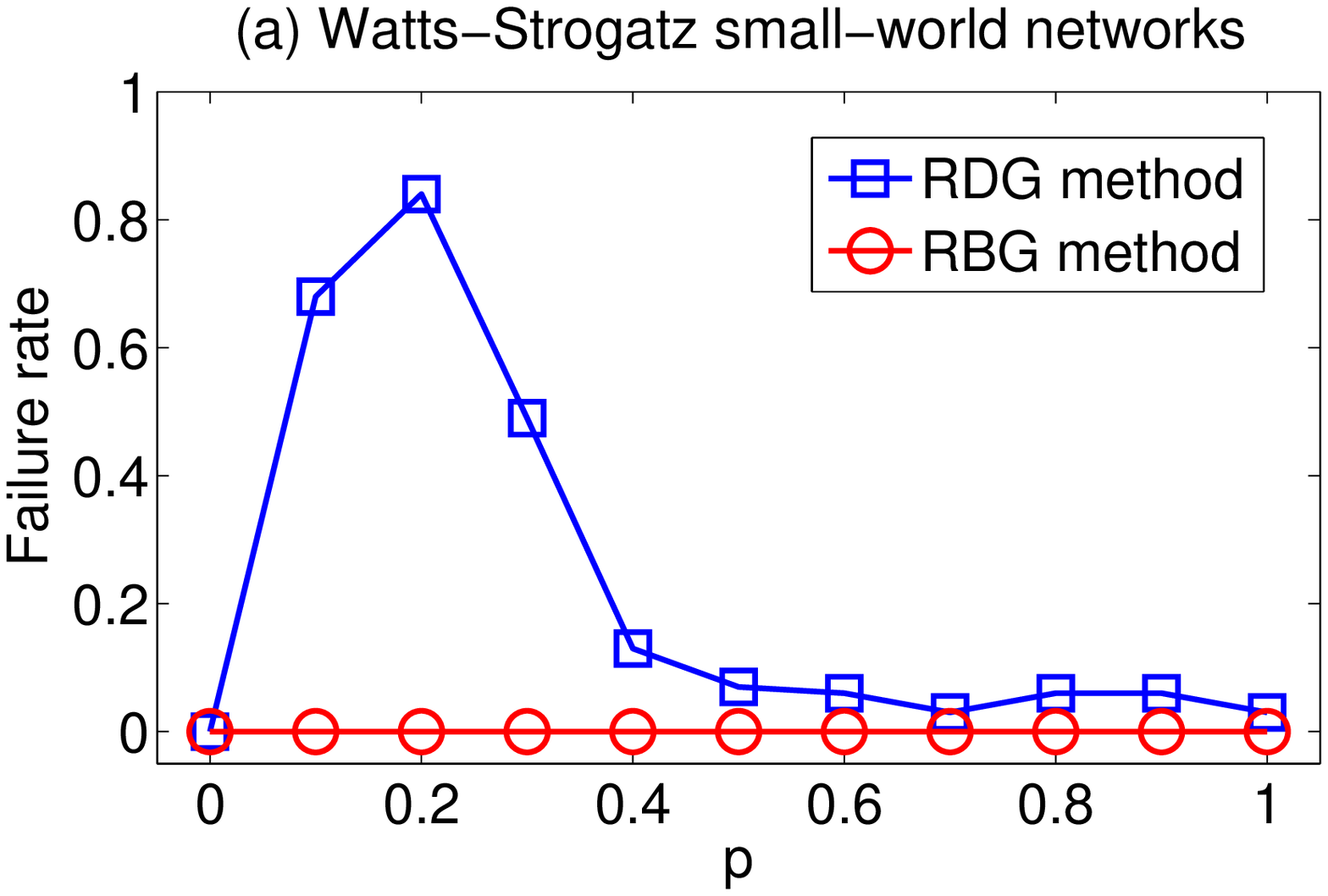}
  \includegraphics[width=4.2cm]{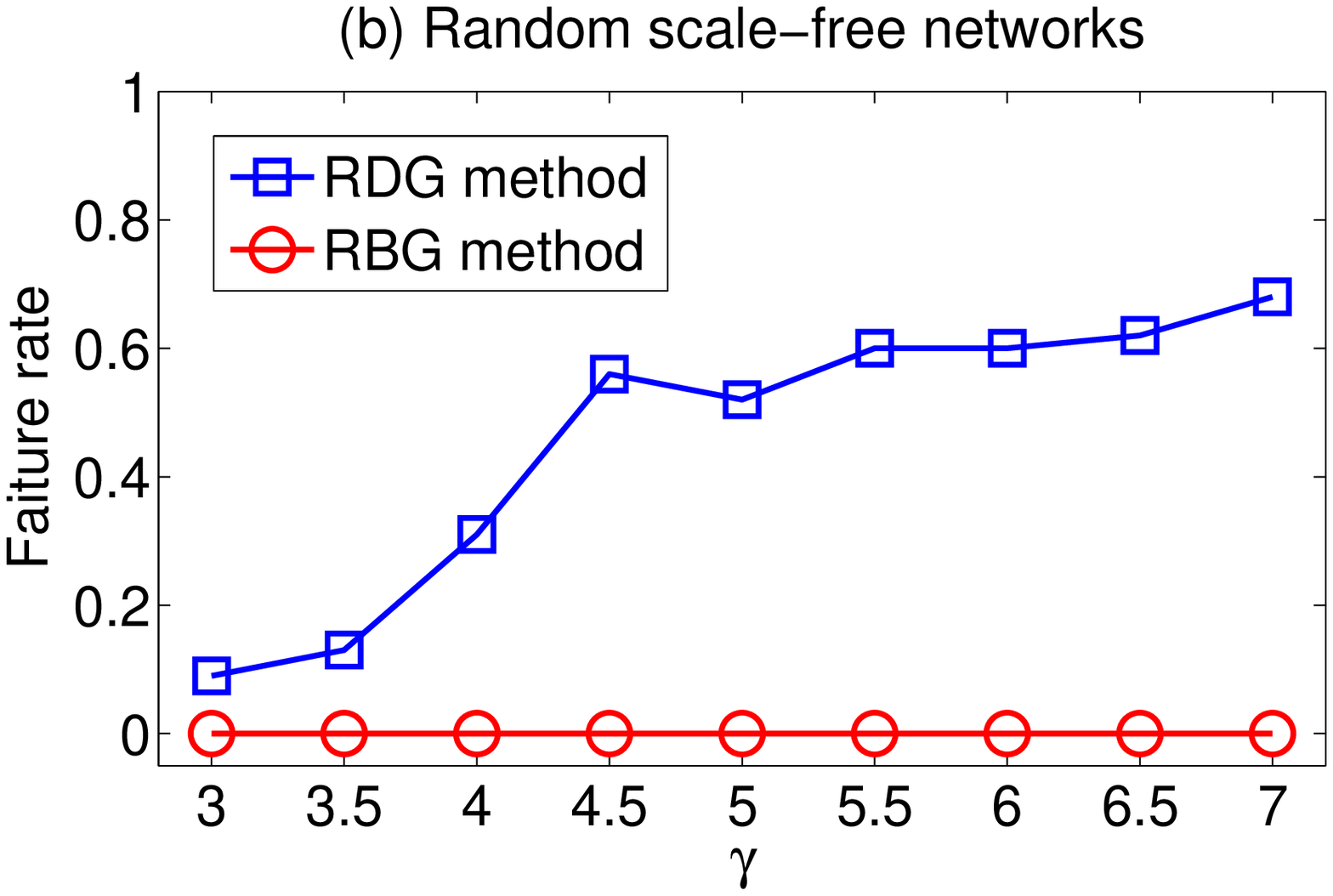}
  \caption{The synchronization failure rate in Watts-Strogatz
small-world networks ($N=500$, $k=6$) and random scale-free networks
($P(k)\thicksim k^{-\gamma}$, $N=500$, $k_{min}=2$) when RDG and RBG
($\theta=0.2$) methods are used. The results are under 100
independent realizations.}
\end{figure}

In conclusion, because RDG method consider only local information
(degree), it will sometimes cause synchronization failure. Based on
the edge betweenness, we claim the RBG method can solve this
problem.


\begin{thebibliography}{99}
\bibitem{explain} Compared with $\lambda^{r}_{min}/\lambda^{r}_{max}$ in
\cite{PRL103}, this index can detect disconnected directed networks
whose $\lambda^{r}_{2}=0$.
\end{thebibliography}
\end{document}